\documentstyle[pra,aps,twocolumn,epsfig]{revtex}

\begin{document}

\title{Continuous optical loading of a Bose--Einstein Condensate}

\author{L. Santos$^{1}$, F. Floegel$^{1}$ ,  T. Pfau$^2$,
and M. Lewenstein$^{1}$}

\address{
(1) Institut f\"ur Theoretische Physik, Universit\"at Hannover,
 D-30167 Hannover,
Germany\\
(2) 5. Physikalisches Institut, Universit\"at Sttutgart, D-70550 Stuttgart}

\maketitle

\begin{abstract}
The continuous pumping of atoms into a Bose--Einstein condensate via
spontaneous emission from a thermal reservoir is analyzed.
We consider the case of atoms with a 
three--level $\Lambda$ scheme, 
in which one of the atomic transitions 
has a very much shorter life--time than the other one.
We found that in such scenario the photon reabsorption in
dense clouds can be considered negligible. If in addition inelastic processes
can be neglected, we find that optical pumping can be used to continuously
load and refill Bose--Einstein condensates, i.e. provides a possible way to
achieve a continuous atom laser. 

\end{abstract}



\section{Introduction}
\label{sec:Intro}

During the last years, the fruitful combination of laser--cooling 
\cite{nobel} and evaporative cooling \cite{BEC} has allowed the experimental
achievement of the Bose--Einstein condensation (BEC)
 in trapped weakly interacting alkali gases. Such remarkable achievement
has stimulated an enormous interest \cite{Varenna}. The
subsequent efforts mainly concentrated on two different areas. 
On one side, the BEC offers an
extraordinary opportunity to test condensed matter and
low--temperature phenomena. In this respect, very recently
several striking results have been reported concerning
superfluidity phenomena \cite{MIT,Oxford} and generation and
dynamics of vortices \cite{JILA,ENS} and dark--solitons
\cite{NIST,Hannover}. On the other hand, 
the macroscopically occupied
matter wave can be manipulated by atom optical elements, that can
be combined to provide new tools for precision experiments
\cite{AtomOptics}. Besides passive optical elements recently also
active elements, that provide phase coherent gain have been
demonstrated \cite{amplifyer}. Also a new
field, called Non--Linear Atom Optics (NLAO), has rapidly
developed during the last years. Several remarkable experiments
have been recently reported in this area, as reflection of BEC
from an optical mirror\cite{Bouncing} and four--wave mixing
\cite{4WM} of matter waves.

Among the results related to BEC and NLAO, one of the most
important experimental achievements was the realization of an {\em Atom
Laser}.  As a
coherent source of matter waves, the
atom laser will lead to new applications in atom optics. Its
impact in the field is comparable to the one of light lasers in
light optics. The first realization of an atom laser was
achieved via pulsed rf-outcoupling from a BEC
\cite{MIToutcoupling}. The coherent character of the source was
demonstrated in a landmark experiment \cite{MITscience}, in which
two atom--laser pulses where overlapped, showing a clear
interference pattern.

Since this first realization, several groups have 
build atom lasers using (quasi-) continuous
outcoupling from the BEC, either by using rf fields
\cite{Munich,Hannoverlaser}, or by employing Raman pulses
\cite{NISTlaser}. However, the continuous outcoupling represents
just a half way towards a cw atom laser. The continuous loading
of the condensate still remains to be incorporated in
experiments. Without a continuous refilling of the BEC, the atom
laser output lasts only as long as some atoms in  the BEC are
kept. Like in the development of light lasers the
availability of cw atom lasers would open the way to "high power"
and precision applications.

Two different physical mechanisms could provide a continuous
pumping into a condensate. On one hand, the collisional
mechanisms \cite{Castin}, in which two non--condensed atoms from
a reservoir collide, and as a result one is pumped into the
condensate, whereas the other carries most of the energy and is
evaporated. On the other hand, the optical pumping of reservoir
atoms into a BEC via spontaneous emission processes has been also
proposed \cite{Martin}. If this reservoir could be filled
in a (quasi-) continuous way  by laser cooling techniques, one
would benefit from the large cooling efficiency of laser cooling
compared to evaporative cooling, allowing for a considerable
increase in atomic flux produced by an atom laser. For the
latter, it is crucial that the spontaneously emitted photons
cannot be reabsorbed, because otherwise a heating is introduced
in the system, and BEC can be neither achieved nor maintained
\cite {Dalibard}.

Several possible dynamical and geometrical solutions 
for the reabsorption problem have been
proposed during the last few years. The geometrical proposals are
based on the reduction of the dimensionality of the traps
\cite{Mlynek,Dalibard}. It is easy to understand that assuming
that the reabsorption cross section for trapped atoms is the same
as in free space, i.e. $\simeq1/k_L^2$, the significance of
reabsorptions increases with the dimensionality, in such a way
that the reabsorptions should not cause any problem in one
dimension, have to be carefully considered in two dimensions,
and  forbid condensation in three dimensions. Therefore,
cigar-shape and disc-shape traps have been suggested. However
even severe deformations of the trap do not allow more than
modest reductions of the reabsorption heating \cite{Reab}. Other
suggestion consists in using a strongly confining trap with a
frequency $\omega\simeq \omega_R$. In this case, it has been
proved \cite{Janicke96} that  in two atom systems the relative
role of reabsorption in such a trap can be significantly reduced.
It is, however, not clear whether this result would hold for many
atom systems. Another promising remedy against reabsorption
heating employs  the dependence of the reabsorption probability
for trapped atoms on the fluorescence rate $\gamma$, which can be
adjusted at will in dark state cooling  \cite{Reab}. In
particular, in the interesting regime in which $\gamma$ is much
smaller  than the trap frequency $\omega$, i.e. in the so called
{\em Festina Lente} limit \cite{Festina}, the reabsorption
processes, in which the atoms change energy and undergo heating,
are practically completely suppressed. However due to the
slow time constants in this approach the cooling efficiency is
greatly reduced. Another reabsorption remedy could be provided
by the destructive quantum interference of the negative effects
of the photon reabsorption in the, so--called,
Bosonic--Accumulation Regime \cite{BAR}.

In this paper, we concentrate on the continuous optical pumping
into a BEC. The paper is divided in two different parts. In the
first part we present a new scenario in which the reabsorption
is suppressed, in much less
restrictive conditions than that of Festina Lente regime, 
i.e. without reducing the cooling efficiency. In this
scenario, an atom possesses an accessible three level $\Lambda$
scheme, in which one of the atomic transitions decays much faster
than the other. 
By employing Master Equation (ME) techniques, we show that 
the dangerous reabsorptions of photons on the slow transition 
are suppressed
because the respective coherences are destroyed due to the 
decay via the fast transition. 
By dangerous we mean here those reabsorptions that may
lead to undesired heating of the system. Since the dangerous
reabsorptions are not present, this scheme can be employed to
continuously pump atoms into the lower level of the slower
transition. In the second part of the paper we analyze the
dynamics of such pumping in the presence of atom--atom
collisions, and show that the combination of elastic collisions
(evaporative cooling), and bosonic enhancement of the spontaneous
emission, can create a condensate, and refill it in the presence
of outcoupling or losses. Therefore, this scheme could be
considered as a possible way towards a continuously loaded atom
laser. As a possible experimental realization we consider
laser cooled Chromium, but the ideas can be generalized to any
atom that provides an asymmetric three--level system.

The structure of the paper is as follows. In Sec.\
\ref{sec:Model} we introduce the physical model, as well as the
quantum ME that determines the loading dynamics. In Sec.\
\ref{sec:BRE} we introduce the so--called Branching Ratio
Expansion (BRE), which allows us to analyze the hierarchy of
processes which occur in the system. In Sec.\ \ref{sec:reab} we
analyze in detail the suppression of the reabsorption effects.
Sec.\ \ref{sec:colis} is devoted to the treatment of the
atom--atom collisions. In Sec.\ \ref{sec:numres} we present the
numerical Monte Carlo results of the loading dynamics. Finally, we
summarize some conclusions in Sec.\ \ref{sec:conclus}.

\section{Model}
\label{sec:Model}

We consider the case of atoms with an accessible
three level $\Lambda$ system (see Fig. 1), formed by the levels
$|r\rangle$, $|e\rangle$ and $|g\rangle$. The atoms are
trapped in an isotropic harmonic trap which, depending of the
internal state of the atoms, has frequencies
$\omega_r$, $\omega_e$ and $\omega_g$, respectively.
This could be, for instance,
 the case of
Chromium $^{52}$Cr, in which the electronic levels would be
$^7$S$_3$,  $^7$P$_4$, and  $^5$D$_4$, respectively. The
transition $|r\rangle\leftrightarrow |e\rangle$ is assumed to be
driven by a laser, which has a Rabi frequency $\Omega$. The
spontaneous emission frequencies associated with the transitions
$|r\rangle\leftrightarrow |e\rangle$ and
$|g\rangle\leftrightarrow |e\rangle$ are, respectively,
$\gamma_{er}$ and $\gamma_{eg}$, such that
$\gamma_{er}\gg\gamma_{eg}$. In the case of $^{52}$Cr,
$\gamma_{er}=2\pi\times 5${\rm MHz} $\gg\gamma_{eg}=2\pi\times
30${\rm Hz}. The branching ratio $\epsilon \equiv
\gamma_{eg}/\gamma_{er}$ is therefore very small ($\sim 10^{-5}$
in $^{52}$Cr). The fact that $\epsilon\ll 1$, will lead to the
suppression of the reabsorption of scattered photons.

\begin{figure}[ht]
\begin{center}\
\epsfxsize=7.5cm
\hspace{0mm}
\psfig{file=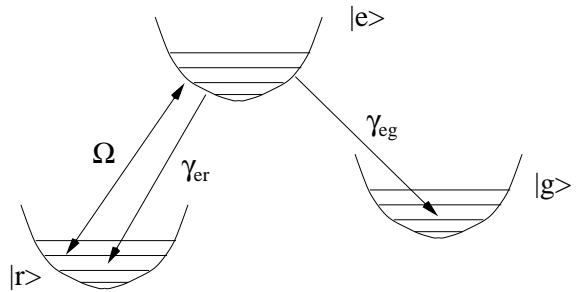,width=7.5cm} \\[0.1cm]
\caption{Atomic scheme considered throughout the paper.}
\label{fig:0}
\end{center}
\end{figure}

In this section, we shall not consider the collisions between the atoms in the
$|g\rangle$ state. Such collisions are introduced in our formalism in
Sec.\ \ref{sec:colis}. In the following we take $\hbar=c=1$
for simplicity. Let us introduce the
annihilation and creation operators of atoms in the $|r\rangle$,
$|e\rangle$ and $|g\rangle$ states and in the trap levels
$s$, $l$, and $m$ which we shall
call $ r_{s},r_{s}^{\dag}$, $ e_{l},e_{l}^{\dag}$, and
$ g_{m},g_{m}^{\dag}$.  These operators fulfill the standard bosonic
commutation relations.

The Hamiltonian which describes the coupling of the system
of bosons to the laser field, as well as
to the vacuum electromagnetic modes is of the form:
\begin{equation}
\hat H= \hat H_0 + \hat H_{er}+\hat H_{af}^{er}+\hat H_{af}^{eg}+\hat H_f,
\label{H}
\end{equation}
with the following terms:
\begin{itemize}
\item Free atomic Hamiltonian (describing internal and
center--of--mass degrees of freedom):
\begin{eqnarray}
&&\hat H_0=\sum_s \omega_s^r r_s^{\dag}r_s+
\sum_l (\omega_l^e+\omega_0)e_l^{\dag}e_l+
\nonumber \\
&&\sum_s \omega_m^g g_m^{\dag}g_m,
\end{eqnarray}
with $\omega_s^r$, $\omega_l^e$ and $\omega_m^g$, denoting the energies of the
level $s$ of the $|r\rangle$ trap, the level $l$ of the $|e\rangle$ trap,
and the level $m$ of the $|g\rangle$ trap, respectively. $\omega_0$
is the transition frequency between
$|r\rangle$ and $|e\rangle$.

\item Interactions of the laser quasiresonant with the transition $|r\rangle\leftrightarrow |e\rangle$:
\begin{equation}
H_{er}=\frac{\Omega}{2}\sum_{l,s}\eta_{l,s}e^{-i\omega_L t}e_l^{\dag}r_s+ h.c.,
\end{equation}
where $\eta_{l,s}$ is the Franck--Condon factor which describes the transition between
a level $s$ of  the $|r\rangle$ trap, and a level $l$ of the $|e\rangle$ trap, and
$\omega_L$ is the frequency of the applied laser.

\item Spontaneous emission processes $|e\rangle\rightarrow |r\rangle$:
\begin{eqnarray}
&&\hat H_{af}^{er}=-i\sum_{l,s}\sum_{\mu}\int d^3\vec k
\sqrt{\frac{k}{2\epsilon_0(2\pi)^3}}(\vec d_{er} \cdot
\vec\epsilon_{\vec k \mu})
\nonumber \\
&&\times\eta_{ls}(\vec k) e_l^{\dag}r_s a_{\vec k \mu} +h.c.,
\end{eqnarray}
where where $a_{\vec k\mu}$, $a_{\vec k\mu}^{\dag}$ are
the annihilation and the  creation
operators of a vacuum photon characterized by a wavevector $\vec k$ and a
polarization $\mu$, with polarization vector $\vec\epsilon_{\vec k\mu}$;
$\vec d_{er}$ is the dipole vector of the transition
$|r\rangle\leftrightarrow |e\rangle$.

\item Spontaneous emission $|e\rangle\rightarrow |g\rangle$:
\begin{eqnarray}
&&\hat H_{af}^{eg}=-i\sum_{l,m}\sum_{\mu}\int d^3\vec k
\sqrt{\frac{k}{2\epsilon_0(2\pi)^3}}(\vec d_{eg} \cdot \vec\epsilon_{\vec k \mu})
\nonumber \\
&&\times \eta_{lm}(\vec k) e_l^{\dag}g_m a_{\vec k \mu} +h.c.,
\end{eqnarray}
where $d_{eg}$ is the dipole vector of the transition
$|g\rangle\leftrightarrow |e\rangle$.

\item Free Hamiltonian of the electromagnetic (EM) field
\begin{equation}
\hat H_f=\sum_{\mu}\int d^3 \vec k k a_{\vec k \mu}.
\end{equation}

\end{itemize}

Starting from the Hamiltonian (\ref{H}),
one can trace the full density matrix of the system
 over the vacuum modes of the EM field.  Using
standard methods of quantum stochastic processes \cite{books}, one can derive
then the Quantum Master Equation in Born--Markov approximation \cite{Santos}.
In principle such ME fully describes all the processes which happen
in the system, including eventual reabsorptions in the fast transition
$|e\rangle\rightarrow |r\rangle$.
However, for simplicity of the analysis we shall consider the case
in which we can neglect the reabsorption phenomena for the reservoir atoms
($|e\rangle$ and $|r\rangle$).
Although what
follows is true also in presence of those reabsorptions, the approximation
 will allow us to
concentrate on  the much simpler  problem
of the pumping of a single atom from
the reservoir $\{|e\rangle,|r\rangle\}$ into the $|g\rangle$ trap, where
of course, collective phenomena are important, and therefore  taken
into account.
In this case, the ME takes the form
\begin{equation}
\dot\rho = -iH_{eff}\rho+i\rho H_{eff}+{\cal J}\rho,
\label{ME}
\end{equation}
where $\rho$ is the density matrix, $H_{eff}=H_{eff}^{(0)}+H_{eff}^{(1)}$ and
${\cal J}={\cal J}^{(0)}+{\cal J}^{(1)}$, with
\begin{mathletters}
\begin{eqnarray}
&&H_{eff}^{(0)}=\sum_s \omega_s^r r_s^{\dag}r_s+ \sum_l (\omega_l^e-\delta
-i\gamma_{er})e_l^{\dag}e_l+ \nonumber \\
&&\sum_s \omega_m^g g_m^{\dag}g_m +H_{er}, \\
&&H_{eff}^{(1)}=-i\gamma_{eg}\sum_{l,m}\sum_{l',m'}
\alpha_{lmm'l'}e_l^{\dag}g_m g_{m'}^{\dag}e_{l'},\\
&&{\cal J}^{(0)}\rho=2\gamma_{er}\sum_{l,s}\sum_{l',s'}
\beta^R_{lss'l'}r_{s'}^{\dag}e_{l'}\rho e_l^{\dag}r_s,\\
&&{\cal J}^{(1)}\rho=2\gamma_{eg}\sum_{l,m}\sum_{l',m'}
\alpha^R_{lmm'l'}g_{m'}^{\dag}e_{l'}\rho e_l^{\dag}g_m.
\end{eqnarray}
\end{mathletters}
Here $\delta=\omega_L-\omega_0$ is the detuning,
\begin{equation}
H_{er}=\frac{\Omega}{2}\sum_{l,s}\eta_{l,s}e_l^{\dag}r_s+ H.c.,
\end{equation}
whereas $\alpha_{lss'l'}=\alpha^R_{lss'l'}+i\alpha^I_{lss'l'}$,
$\beta_{lmm'l'}=\beta^R_{lmm'l'}+i\beta^I_{lmm'l'}$, with
\begin{mathletters}
\begin{eqnarray}
&&\alpha^R_{lmm'l'}=\int d\hat k {\cal W}(\hat k)
\eta_{ls}(k_0^{eg}\hat k)\eta_{l'm'}^{\ast}(k_0^{eg}\hat k),\\
&&\alpha^I_{lmm'l'}=\frac{-1}{\pi} P \int_{-\infty}^{\infty}d\xi
\frac{\xi^3}{\xi-1}
\int d\hat k {\cal W}(\hat k) \nonumber \\
&&\times \eta_{lm}(\xi k_0^{eg}\hat k)\eta_{l'm'}^{\ast}(\xi k_0^{eg}\hat k)).
\end{eqnarray}
\end{mathletters}
In the previous equations $\hat k$ indicates the solid angle
coordinates, ${\cal W}(\hat k)$ represents the dipole pattern,
$k_0^{eg}$ is the wave number associated with the transition
$|g\rangle\leftrightarrow |e\rangle$, and $P$ denotes the Cauchy
Principal part. Finally, $\beta^R_{lss'l'}$ 
has an identical form as $\alpha^R_{lmm'l'}$,
but for the transition $|r\rangle\leftrightarrow |e\rangle$.

\section{Branching Ratio Expansion}
\label{sec:BRE}

In this section we shall show that if $\epsilon\ll 1$, the
reabsorption effects on the slow transition
can be safely neglected, because they occur 
with a probability $1/\epsilon$ times
smaller than the spontaneous emission processes
$|e\rangle\rightarrow |g\rangle$ without any reabsorption. In
order to demonstrate that, we shall perform the analysis of the
different processes that
 could happen in the system.

We shall consider the situation in which 
some atoms are already accumulated in the level $|g\rangle$ and 
the time $t_0$ a single atom is being pumped from the level 
$|r\rangle$ to the level $|e\rangle$. This atom undergoes then 
the spontaneous emission process, and its further fate may be 
twofold. First, it may undergo transition 
$|e\rangle\rightarrow |r\rangle$, and the emitted photon will 
then leave the system (we assume no reabsorptions at 
$|e\rangle\leftrightarrow |r\rangle$ line due to the low density 
of $|r\rangle$ atoms). Second, the excited atom may undergo 
a transition $|e\rangle\rightarrow |g\rangle$. We allow for the 
possibility that the emitted photon in this process may be 
reabsorbed several times, until it leaves the system or until 
the emission $|e\rangle\rightarrow |r\rangle$ will take place.
We shall analyze the hierarchy of possible processes which 
can be produced on the
time scale of the above described processes. 
We shall assume that no other atom is pumped from the level 
$|r\rangle$ to $|e\rangle$ within this time scale.
Technically, this approximation
means that we exclude the possibility of multiple quantum
jumps from $|e\rangle$ to $|r\rangle$, i.e. we assume that the
whole process consists of the sequence of processes, each
involving an atom being pumped from $|r\rangle$ to $|e\rangle$,
which then undergoes spontaneous emission processes (including
reabsorptions) until it either lands on the level $|r\rangle$ or
$|g\rangle$, after which another atom is being pumped from
$|r\rangle$ to $|e\rangle$, and so on. 

This approximation is
 performed here just for reasons of technical simplicity, but
we want to stress that:
\begin{itemize}
\item The approximation describes
well the experimental situation with Chromium atoms
in which the Rabi frequency $\Omega<\gamma_{er}$.
The pumping process has thus indeed an incoherent character, 
consisting in a sequence of jumps $|e\rangle\rightarrow |r\rangle$
followed by spontaneous emission events. Of course, 
it may happen that several atoms are being excited 
simultaneously to $|e\rangle$. Each of them, however, 
behaves independently of the others, so that the analysis 
pertaining to just one excited atom is valid. 
\item The result obtained below
is indeed more general, because at any time scale the hierarchy of
probabilities is maintained. The latter statement means that our results 
hold also for $\Omega>\gamma_{er}$. This can be at best understood using 
a dressed--state picture with respect to the  
$|e\rangle\leftrightarrow |r\rangle$ transition.
In the limit $\Omega\gg\gamma_{er}$ the dynamics reduces to 
a situation in which a single atom is being pumped to one 
of the dressed states $|+\rangle$ ($|-\rangle$), and undergoes spontaneous 
emission consisting in 
arbitrary number of incoherent
$|+\rangle\rightarrow|+\rangle$ 
($|-\rangle\rightarrow|-\rangle$) transitions, 
ending either in $|g\rangle$, or in $|-\rangle$ ($|+\rangle$). 
This process is followed by an arrival of another atom in the state 
$|+\rangle$ ($|-\rangle$), followed by successful spontaneous emission, etc.
Each one of these steps can be understood with the model presented below.
\end{itemize}

The formal solution of the ME (\ref{ME}) after
a photon escapes from the system is given by:
\begin{equation}
\rho_{\infty}=\int_0^{\infty}dt {\cal J} \{ e^{-i{\cal L} _{eff}t}\rho \}.
\label{inf}
\end{equation}
We have used in the expression (\ref{inf}) the shortened
notation ${\cal L}_{eff}\rho=-iH_{eff}\rho+i\rho H_{eff}^{\dag}$.
In the following we shall denote
${\cal L}_{eff}^{(0)}\rho=-iH_{eff}^{(0)}\rho+i\rho H_{eff}^{(0)}$,
${\cal L}_{eff}^{(1)}\rho=-iH_{eff}^{(1)}\rho+i\rho H_{eff}^{(1)}$.
Since $\gamma_{eg}\ll\gamma_{er},\Omega$, we are going to perform an expansion in the
branching ratio $\epsilon$. We consider an initial state of the system given by
$|\psi_0\rangle=|\psi_0^g\rangle\otimes|\psi_0^{e,r}\rangle$, with
$|\psi_0^g\rangle=|N_0,N_1,\dots,N_m,\dots\rangle$, with $N_m$ denoting
the initial population
of the $m$-th state of the $|g\rangle$ trap. $|\psi_0^{e,r}\rangle$ denotes
the initial state of the $|r\rangle$ and $|e\rangle$ traps.
We are interested
in the probability to obtain a final state
$|\psi_f\rangle=|\psi_f^g\rangle\otimes|\psi_f^{e,r}\rangle$.

The branching ratio expansion must take into account the bosonic
enhancement effects.  Due to the bosonic enhancement, and the
fact that we consider that the ground state of the $|g\rangle$
can be macroscopically populated, the expansion must be done in
the parameter $\epsilon N_0$, and not only in $\epsilon$.
Nevertheless, we expect that the expansion and the conclusions
that we draw from it will remain valid until $\epsilon N_0\simeq
1/2$. To use this expansion
 in the considered experimental situation
we have to limit ourselves to  the case of $N_0\le 10^5/2$.
That means, however,  that the expansion can nevertheless
be safely used  during the
onset of the condensation, and the dangerous reabsorbtions can be
neglected in that regime.
The situation in which $\epsilon N_0 \agt 1$ requires the treatment
of higher order terms in the expansion, but does not mean that
in such a case the reabsorbtions will  cause troubles.
In the case when very many atoms are already condensed, the dynamics is dominated by bosonic statistical effects. The
use of similar ideas and
techniques as those employed in the Boson Accumulation Regime (BAR)
\cite{BAR} should be then possible.

Employing standard methods of time--dependent perturbation 
theory in the small parameter $\epsilon N_0$
we obtain:
\begin{equation}
\langle \psi_f |\rho_{\infty}|\psi_f\rangle= A_0+A_1+A_2+{\cal O}((\epsilon N_0)^3),
\end{equation}
where $A_j$ is the term of order ${\cal O}((\epsilon N_0)^j)$. Let us
now analyze step by step each of the
terms of the branching ratio expansion (BRE):

\subsection{Zeroth Order}
\label{sec:Ord0}

The zeroth order term is of the form:
\begin{equation}
A_0=\langle \psi_f |\int_0^{\infty}dt
{\cal J}^{(0)}\{ e^{{\cal L}_{eff}^{(0)}t}\rho_0 \}|\psi_f\rangle,
\label{A0}
\end{equation}

This process is by far the most probable one, and implies
an spontaneous emission into the $|r\rangle$ state 
These processes (and the subsequent repumping by the laser) 
induce a thermal distribution of the
$|e\rangle$ and $|r\rangle$ traps, and do not affect directly the populations
of the $|g\rangle$ trap.

\subsection{First Order}
\label{sec:Ord1}

\begin{figure}[ht]
\begin{center}\
\epsfxsize=5.0cm
\hspace{0mm}
\psfig{file=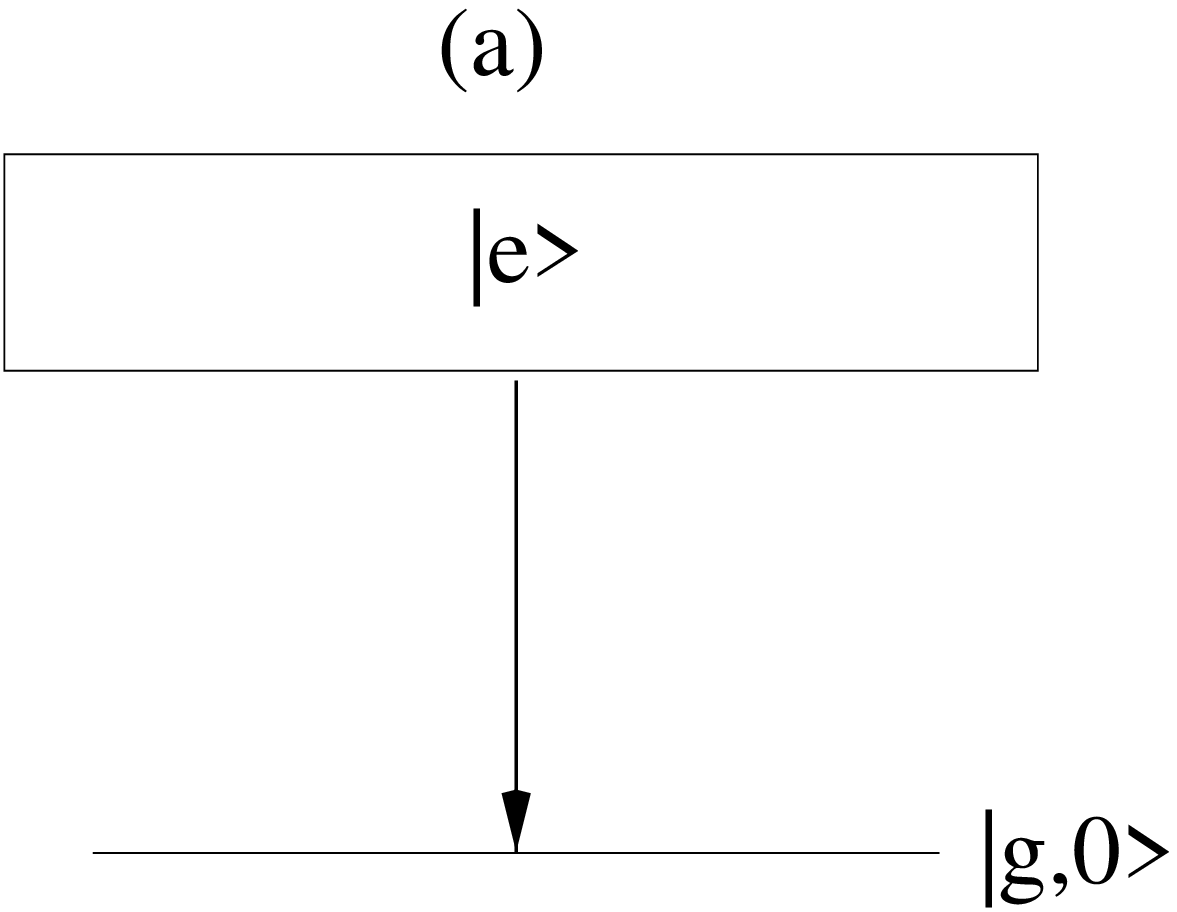,width=4.0cm} \psfig{file=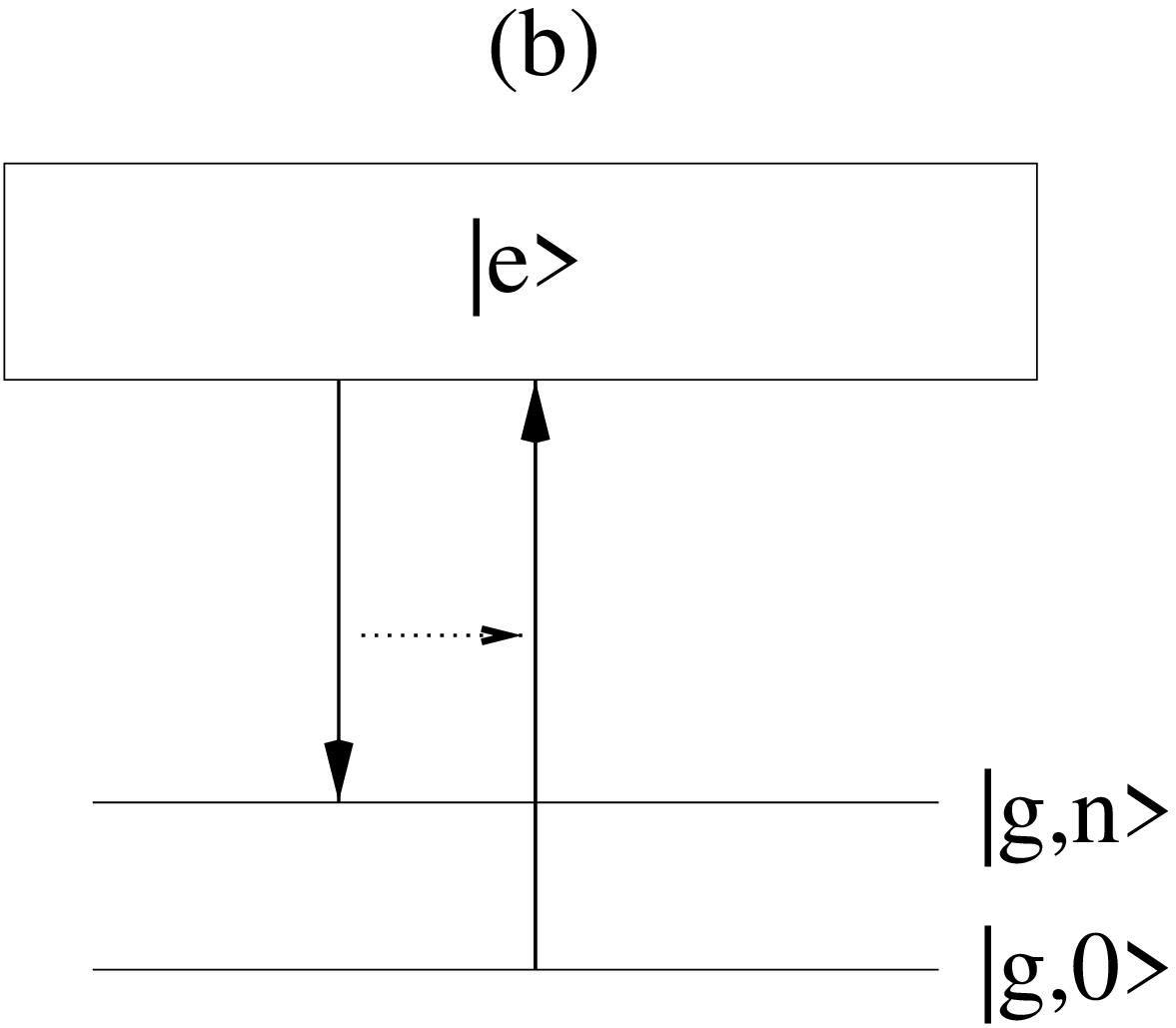,width=4.0cm}\\[0.2cm]
\psfig{file=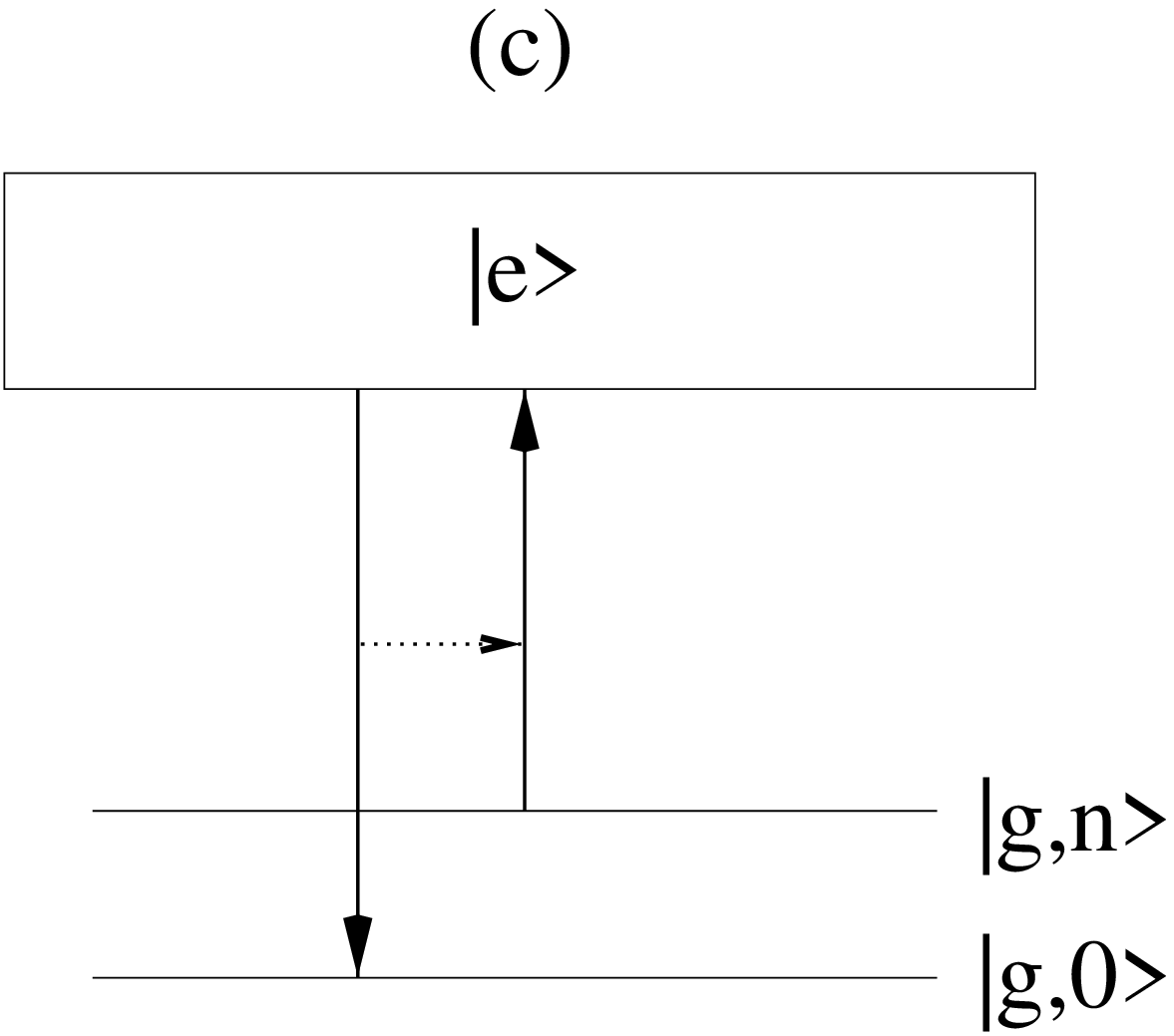,width=4.0cm} \psfig{file=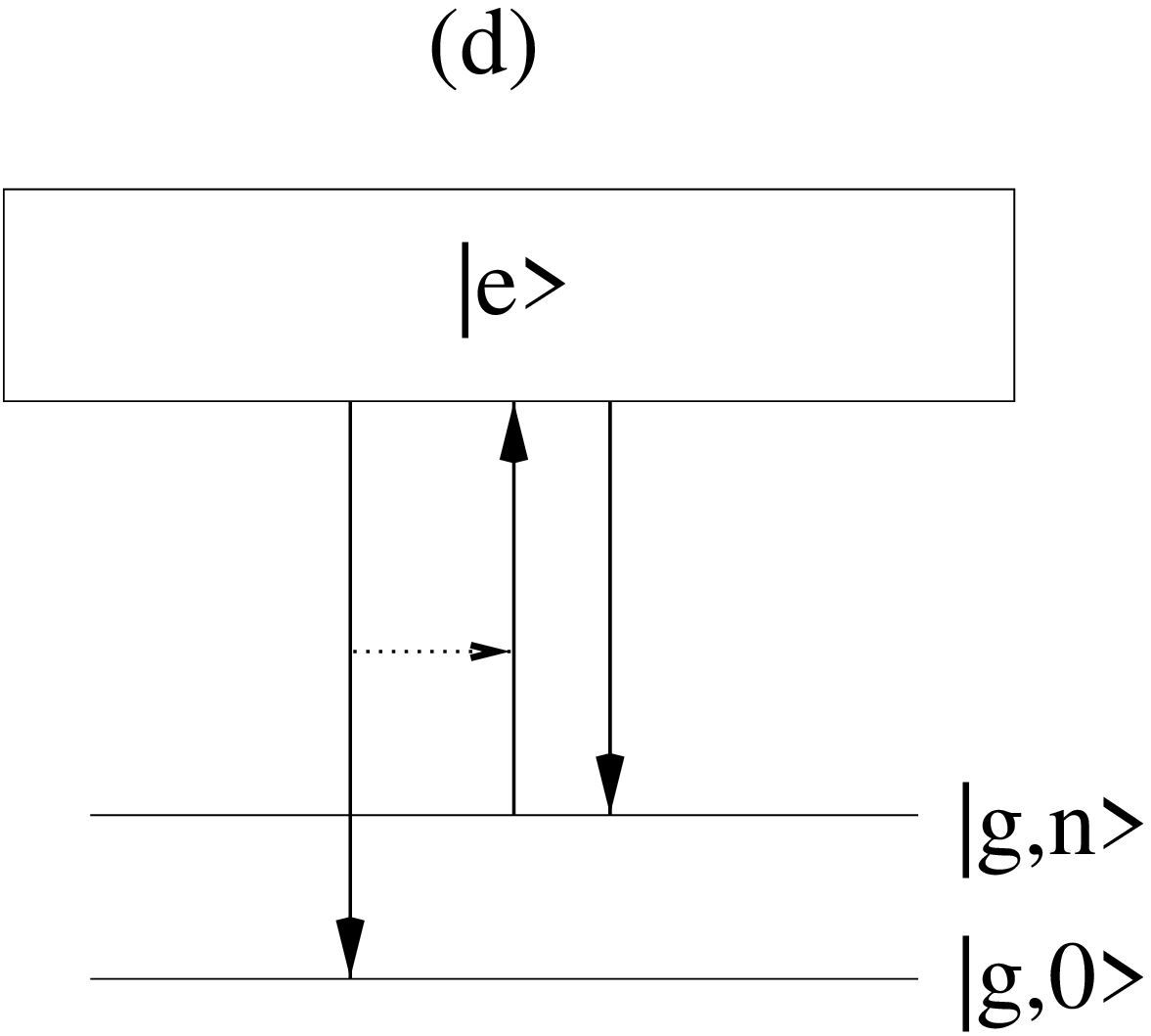,width=4.0cm}\\[0.2cm]
\psfig{file=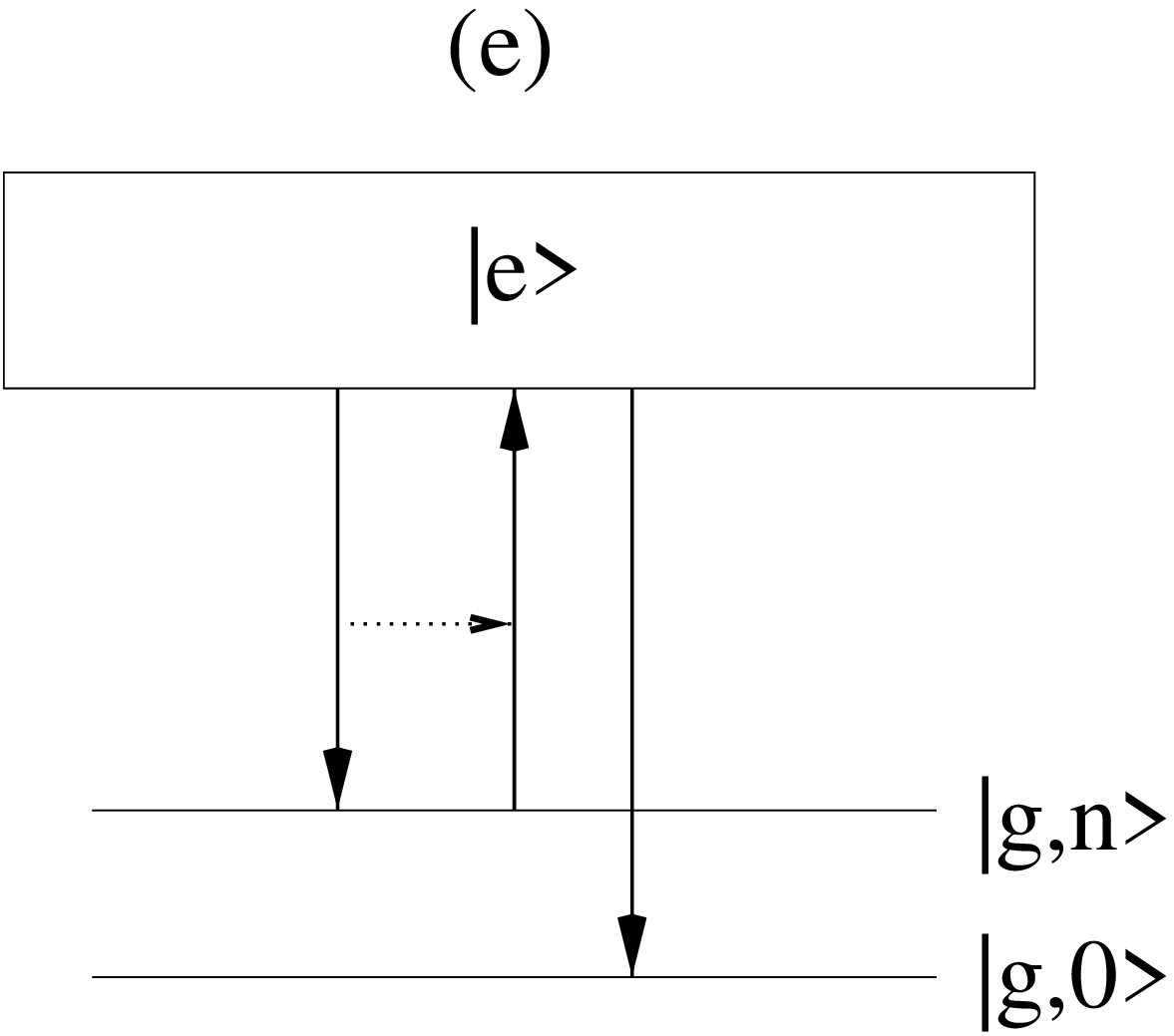,width=4.0cm}\\[0.2cm]
\caption{Scheme of different processes which appear in the BRE.
The solid line implies decays or reabsorptions, whereas the
dashed line represents the photon which is spontaneously
emitted and further reabsorbed.
(a) represents the process of first BRE order $A_{1a}$;
(b) and (c) represent respectively the negative and positive 
reabsorption effects which appear in the second BRE order processes 
$A_{2a}$; the interference between the processes (a), and (d,e)
constitutes the term $A_{2b}$ of the second order of the BRE.}
\label{fig:1}
\end{center}
\end{figure}

The first order term of the BRE is of the form $A_1=A_{1a}+A_{1b}$, where
\begin{eqnarray}
&&A_{1a}=\langle \psi_f |\int_0^{\infty}dt
{\cal J}^{(1)}\{ e^{{\cal L}_{eff}^{(0)}t}\rho_0 \}|\psi_f\rangle,\label{A1a} \\
&&A_{1b}= \langle \psi_f |\int_0^{\infty}dt
{\cal J}^{(0)} \nonumber \\
&& \left\{ \int_0^{t}dt'e^{{\cal L}_{eff}^{(0)}(t-t')}
\left\{ {\cal L}_{eff}^{(1)} \left\{ e^{{\cal L}_{eff}^{(0)}t'}\rho_0
\right\} \right\} \right\} |\psi_f \rangle. \label{A1b}
\end{eqnarray}
$A_{1a}$ corresponds to the case in which an
atom in the state $|e\rangle$ decays (without any further reabsorption) into a
state of the $|g\rangle$ trap (see Fig.\ \ref{fig:1}(a)).
$A_{1b}$ is given by the interference of two
different process: a process like the one considered in the zeroth order, and
a process in which (i) a decay is produced into some state of the
$|g\rangle$ trap,
(ii) a subsequent reabsorption is produced from the same state of the
$|g\rangle$ trap,
and finally a process as that of the zeroth order is produced.
The processes described
by $A_{1b}$, although containing reabsorptions,
 do not change the population distribution of the
$|g\rangle$ trap, and
can be considered as small quantitative corrections to the zeroth order term.

\subsection{Second Order}
\label{sec:Ord2}

The second order term of the BRE is of the form
$A_2=A_{2a}+A_{2b}$, with
\begin{mathletters}
\begin{eqnarray}
&& A_{2a}=
\langle \psi_f |\int_0^{\infty}dt
{\cal J}^{(0)} \nonumber \\
&&\left\{ \int_0^{t}dt'e^{{\cal L}_{eff}^{(0)}(t-t')}
\left\{ {\cal L}_{eff}^{(1)} \left\{
\right\delimiter 0 \right\delimiter 0 \right\delimiter 0 \nonumber \\
&& \left\delimiter 0 \left\delimiter 0 \left\delimiter 0
\int_0^{t'}dt''e^{{\cal L}_{eff}^{(0)}(t'-t'')}
\left\{ {\cal L}_{eff}^{(1)} \left\{ e^{{\cal L}_{eff}^{(0)}t'}\rho_0
\right\} \right\} \right\} \right\} \right\} \label{A2a} \\
&&A_{2b}=
\langle \psi_f |\int_0^{\infty}dt
{\cal J}^{(1)} \nonumber \\
&&\left\{ \int_0^{t}dt'e^{{\cal L}_{eff}^{(0)}(t-t')}
\left\{ {\cal L}_{eff}^{(1)} \left\{ e^{{\cal L}_{eff}^{(0)}t'}\rho_0
\right\} \right\} \right\}. \label{A2b}
\end{eqnarray}
\end{mathletters}
Let us consider the term $A_{2a}$. This term involves processes in which an
atom originally in some state of the $|e\rangle$ trap, decays into some
state $m$ of the $|g\rangle$ trap, producing an spontaneously emitted photon,
which is reabsorbed by other atom in some other state $n\not= m$
of the $|g\rangle$ trap. These processes are of order $\epsilon^2$, except
the case in which $n=0$ or $m=0$ (Figs.\ \ref{fig:1}(b) and (c));
in such a case, if the system is already
condensed, the probability associated with these processes is of order
$\epsilon^2 N_0$. We must note that the process of Fig.\ \ref{fig:1}(b)
introduces a negative effect of the photon reabsorption
in our system, because produces a non--condensed atom, while destroys
an already condensed one (of course the opposite process of
Fig.\ \ref{fig:1}(c) corresponds to positive effects, and is
of the same order).
The term $A_{2b}$ is due to the interference effects between the process
considered in $A_{1a}$ and the processes of  Figs.\ \ref{fig:1}(d) and (e).
These process do not cause any negative or positive effects
of the reabsorption,
and simply introduce small quantitative corrections to $A_{1a}$.

As observed, the ``bad'' reabsorption processes which change the
$|g\rangle$ trap population distribution (i.e. may lead to
undesired heating) are of order $\epsilon^2N_0$, and therefore
are $1/\epsilon$ times less probable than the single spontaneous
emission into the $|g\rangle$ trap without any reabsorption.
Hence, the reabsorption effects can be safely neglected, i.e. the
atoms in the $|g\rangle$ trap can be considered as transparent for the
spontaneously emitted photons on the $|e\rangle\rightarrow
|g\rangle$ transition. We shall show in Sec.\ \ref{sec:numres}
that this effect can be used to optically pump atoms from the
reservoir $\{ |r\rangle , |e\rangle \}$ into the $|g\rangle$
trap, and eventually into a condensate created in it.

\section{Suppression of the reabsorption effects}
\label{sec:reab}

In this section we analyze in detail the physical effect behind
the decreasing of the reabsorption probability in the considered system.
The physical picture can be understood by taking a closer look to the
expression (\ref{A2a}) which, after eliminating the terms which do
not change the populations of the $|g\rangle$ trap, takes the form
\begin{eqnarray}
&&\int_0^{\infty}dt\int_0^{t}dt'\int_0^{t'}dt''
\langle\psi_f|{\cal J}^{(0)} \nonumber \\
&&\left\{
|\psi (t;t')\rangle\langle\psi (t;t'')|+
|\psi (t;t'')\rangle\langle\psi (t;t')|)
\right\}|\psi_f\rangle,
\label{A2a2}
\end{eqnarray}
where
\begin{equation}
|\psi (t;t')\rangle=e^{-iH_{eff}^{(0)}(t-t')}H_{eff}^{(1)}
e^{-iH_{eff}^{(0)}t'}|\psi_0\rangle.
\end{equation}
Therefore, the process can be divided into three parts:
(i) From time $0$ to some $t'$, the system evolves following the
effective Hamiltonian $H_{eff}^{(0)}$;
(ii) At $t'$ a spontaneous emission occurs from $|e\rangle$
to $|g\rangle$, followed by a reabsorption;
(iii) The system undergoes after $t'$ the same dynamics as
in the part (i), until time $t$ where a jump ${\cal J}^{(0)}$ is
produced into the $|r\rangle$ state.

As observed in the expression (\ref{A2a2}), the term $A_{2a}$ depends
on the correlation of the amplitudes of probability of
two processes (i--iii) in which (ii)
is produced at two different times, $t'$ and $t''<t'$.
In the interval $t''$ to $t'$ a jump into $|r\rangle$
is  produced with large probability, and therefore the probability
amplitude is reduced, roughly speaking,
by a factor $\exp (-\gamma_{er}(t'-t''))$.
Therefore, the correlation decays very rapidly (in comparison with
$\gamma_{eg}^{-1}$)
with the time difference $t'-t''$. This leads to the strong reduction
of the reabsorption probability.

A very intuitive picture of the underlying physics in
3-level atoms can be obtained in the following way: The
reabsorption on the $|g\rangle$ to $|e\rangle$ transition has an
oscillator strength $\gamma_{eg}$. However the excited state
coherence is rapidly destroyed due to the decay of state
$|e\rangle$ to $|r\rangle$, with a rate $\gamma_{er}$. Therefore
the absorption line width on the $|g\rangle$ to $|e\rangle$
transition is dominated by $\gamma_{er}$. Similar to other
broadening mechanisms the effective reabsorption cross section is
reduced by a factor $\epsilon = \gamma_{er}$/$\gamma_{eg}$. The
atomic sample can be $\epsilon$-fold more dense than in the
normal two level case before it becomes optically thick and
reabsorption becomes a relevant process.

It is perhaps interesting to mention here
 the differences and similarities with the
Festina Lente regime \cite{Festina}, which constitutes a known
way to avoid the reabsorption problem. As pointed out previously,
the reabsorption probability is determined (as already shown in
\cite{Festina}) by the correlation at different times of the
amplitude of reabsorption probability. In the case considered in
\cite{Festina}, the correlation decays with the same frequency
$\Gamma$ as the spontaneous emission frequency of the system, and
therefore the probability of decay plus reabsorption is that of
the decay without further reabsorption (multiplied by some
geometrical factor, specially important in the so--called
free--space limit \cite{Festina}). In the case of the Festina
Lente regime ($\Gamma\ll\omega$) \cite{Festina} the reabsorptions
which do not preserve the energy are suppressed due to a
different reason than that considered in the present paper. In the
Festina Lente conditions the interference terms in the amplitude
of probability of ``bad'' reabsorptions at different times have a
phase which rapidly oscillates in time; therefore, the time
integration leads to a strong reduction of the reabsorption
probability (by the small factor $\Gamma/\omega$). In this sense,
therefore, the Festina Lente Regime can be understood in terms of
an expansion similar to the BRE. The case considered in this
paper is, however, different. The spontaneous emission rate
 $\gamma_{eg}$ is now a factor $\epsilon$
smaller than the rate of decay of the
correlation, given by $\gamma_{er}$. This explains why the
reabsorption is a factor $\epsilon$ less probable than a decay
without reabsorption.

As a final remark, let us point out that the BRE does not
necessarly imply small spontaneous emission rates
($\gamma_{eg}<\omega$), as it was the case of Festina Lente,
but just the relative condition $\gamma_{eg}\ll\gamma_{er}$. In
particular, $\gamma_{eg}$ could be larger (or even much larger)
than the trap frequency. Therefore, in principle, the atoms could
be pumped into the $|g\rangle$ trap much faster than in the Festina
Lente limit. This is of special
interest when considering a sufficiently fast continuous loading
of a condensate.

\section{Collisions}
\label{sec:colis}

In this section we introduce the collisions between atoms in the
$|g\rangle$ trap. In particular, no collision is considered between the atoms
in the $|g\rangle$ trap, and those in the $|e\rangle$ and $|r\rangle$ trap.
Such approximation is valid assuming the situation
in which the reservoir atoms are at much larger temperature
and lower densities than those atoms in the $|g\rangle$ trap.
Since we shall consider a $|g\rangle$ trap of a
finite depth, the eventual collisions with
the relatively much hotter reservoir atoms would lead to
losses in the $|g\rangle$ trap, and eventually to
a depopulation of the condensate created in it.
Since we consider a loss mechanism from the condensate
anyway (Sec.\ \ref{sec:numres}) such collisional losses
could be easily taken into account phenomenologically in our
simulations as an effective outcoupling rate.

The effects of the elastic collisions within the $|g\rangle$ trap
are accounted by a new term in the Hamiltonian
(\ref{H}):
\begin{equation}
\hat H_{coll}=\sum_{m_1,m_2,m_3,m_4}\frac{1}{2}U_{m_1,m_2,m_3,m_4}
g_{m_4}^{\dag}g_{m_3}^{\dag}g_{m_2}g_{m_1}.
\label{Hcoll}
\end{equation}
In the regime we want to study, only the $s$--wave scattering is important,
and then:
\begin{equation}
U_{m_1,m_2,m_3,m_4}=\frac{4\pi\hbar^2a}{m}\int_{R^3}d^3x
\ \psi_{m_4}^{\ast}\psi_{m_3}^{\ast}\psi_{m_2}\psi_{m_1},
\end{equation}
where $\psi_{m_j}$ denotes the harmonic oscillator wavefunctions and
$a$ denotes the scattering length.

In the following we are going to work in the so--called
{\em weak--condensation regime}, where  no
mean--field effects are considered. This means that we consider that
the mean--field energy provided by
the collisions is smaller than the oscillator energy.
Therefore the model is only valid to describe either the onset of
the condensation, or the full condensation
but with the constraint that
the condensate cannot
contain more than, say 500--1000 particles.
A realistic calculation of the dynamics
beyond the weak--condensation regime
would require the self--consistent calculation of the
atomic states, which due to mean--field effects would change
during the dynamics. Such calculation could be possible
by using Bogolyubov--de Gennes formalism, and will be the
subject of further investigation.
In this paper we concentrate in the weak--condensation regime,
where one can apply the formalism of Quantum--Boltzmann Master Equation
(QBME) \cite{QK1,QK2} to treat the collisional effects.
By using similar arguments as those employed in the context of
collective laser cooling in the
presence of atomic collisions in the weak--condensation
regime \cite{colis}, one can show that the ME which describes
in this regime
the loading dynamics from the reservoir $\{ |e\rangle ,|r\rangle \}$
to the $|g\rangle$ can be divided into two independent parts:
\begin{equation}
\dot \rho(t)={\cal L}_{coll}\rho(t) + {\cal L}_{load}\rho(t)
\label{ME2}
\end{equation}
where ${\cal L}_{load}\rho(t)$ is the rhs of Eq. (\ref{ME}), 
and ${\cal L}_{coll}$
describes the collisions, and has the form of a QBME as that described
in Refs. \cite{QK1,QK2}.

Summarizing, the dynamics of the system splits into two parts, (i) collisional part,
described by a QBME, and (ii) loading part, described by the same
ME (\ref{ME}) as without collisions.  The independence of both dynamics,
constitutes the main technical advantage of considering the weak--condensation
regime, and
allows for an easy simulation of the loading process
in the presence of $|g\rangle$--$|g\rangle$ collisions. In particular, we simulate both dynamics using Monte Carlo methods,
combining the numerical method of Ref.\ \cite{QK2}, with simulations similar
as those already presented in Refs.\ \cite{Santos}.

The numerical simulation of the collisional dynamics for a
three--dimensional harmonic trap is demanding, 
due to both the degeneracy of the
levels, and the difficulties to obtain reliable values for the
integrals $U(n_1,n_2,n_3,n_4)$. Therefore, we shall limit
ourselves to the use of the ergodic approximation
\cite{QK2,Holland}, i.e. we shall assume that states with the
same energy are equally populated. The populations of the
degenerate energy levels equalize on a time scale much faster
than the collisions between levels of different energies, and
than the loading typical time. Following Ref.\ \cite{Holland} the
probability of a collision of two atoms in energy shells $m_1$
and $m_2$, to give two atoms in shells $m_3$ and $m_4$ (where
this collision is assumed to change the energy distribution
function), is of the form:
\begin{eqnarray}
&&P(m_1,m_2\rightarrow m_3,m_4)=\Delta (m_j+1)(m_j+2)\times \nonumber \\
&&\frac{N_{m_1}(N_{m_2}-\delta_{m_2,m_1})(N_{m_3}+g_{m_3})(N_{m_4}+g_{m_4}+\delta_{m_3,m_4})}
{g_{m_1}g_{m_2}g_{m_3}g_{m_4}},
\end{eqnarray}
where $g_{m_k}=(m_k+1)(m_k+2)/2$ is the degeneracy of the energy shell $m_k$,
$m_j={\rm min} \{ m_1,m_2,m_3,m_4 \}$,
and $\Delta=(4a^2\omega_g^2m)/(\pi\hbar)$.
In the following we shall use for the calculations
the mass of $^{52}$Cr, and, since $a$ is not known for $^{52}$Cr,  we
shall adopt  a scattering length $a=6$ nm (similar
to that of Rubidium). We shall consider a trap frequency $\omega_g=2\pi\times
1${\rm kHz}.

\section{Numerical Results}
\label{sec:numres}

In this section we consider two different physical problems. First, we shall
show that under appropriate conditions it is possible to load an
initially empty trap for atoms in the state $|g\rangle$
via spontaneous emission from a thermal
reservoir of particles $|e\rangle$. This allows to achieve the condensation
in the trap $|g\rangle$ in a finite time, which depends on the
physical parameters. Secondly, we shall show that in presence
of an outcoupling (or, as discussed in Sec.\ \ref{sec:colis}, in presence
 of trap losses) it is possible to maintain the condensate population
by refilling the condensate via the spontaneous emission from the
thermal cloud.

The zero order term of the BRE corresponds to the 
fastest transition, but does not affect directly the loading
process. Here we shall concentrate on the process provided by
$A_{1a}$, i.e. the spontaneous emission $|e\rangle\rightarrow
|g\rangle$. The loading rate of a state $|g,n\rangle$ from a state
$|e,l\rangle$ is therefore given by
\begin{equation}
\Gamma(l,n)=2\gamma_{eg}\int d\hat k {\cal W}(\hat k)
|\langle n|e^{i\vec k\vec r}|l\rangle|^2(N_n+1).
\end{equation}
Let us assume that $N_{ex}$ atoms are in the level $|e\rangle$ with a thermal
distribution of temperature $T$. The total loading rate into the state
$|g,n\rangle$ is provided by:
\begin{equation}
\Gamma(n)=\sum_l p(l) \Gamma(l,n),
\end{equation}
where $p(l)=N_{ex}\exp\{-\hbar\omega_e\tilde l/k_BT\}/{\cal Z}$ is the
thermal distribution, with  $\tilde l=l_x+l_y+l_z$ and
${\cal Z}=(1-\exp\{-\hbar\omega_e/k_BT\})^{-3}$
is the canonical partition function.
Thus, the loading rate can be rewritten as:
\begin{eqnarray}
&&\Gamma(n)=2\gamma_{eg}\int d\hat k {\cal W}(\hat k) \nonumber \\
&&\langle n|e^{i\vec k\vec r}
\left [ \sum_l N_{ex} \frac{e^{-\hbar\omega_e\tilde l/k_BT}}{{\cal Z}}
|l\rangle\langle l| \right ]
e^{-i\vec k\vec r}|n\rangle (N_n+1)
\label{gamn}
\end{eqnarray}

Let us consider now the following conditions which greatly simplify
the numerical calculation of the loading process.
We are going to assume that $\omega_e<\omega_g+2\omega_{rec}/3$, so
that there is always at least a state of the excited--state
trap in an interval
$\pm\omega_{rec}$ (with $\omega_{rec}$ being the recoil frequency)
around any considered state of the ground--state trap.
The ground state trap has a finite depth, i.e. it possesses a
maximal energy shell at $\hbar\omega_g m_{max}$ \cite{foot2}.
Therefore, only those
$|e\rangle$ atoms with energies $E<\hbar\omega_g m_{max}+\hbar\omega_{rec}
\equiv E_{max}$ can effectively
load the trap. Let us consider that the reservoir has a temperature
$T$ such that $k_BT\gg E_{max}$ \cite{foot2}.
Under the previous conditions, the expression between the brackets
in (\ref{gamn}) can be safely substituted by $1/{\cal Z}$.
Since the previous conditions also imply
$\hbar\omega_e\ll k_BT$, the loading rate can be rewritten as:
\begin{equation}
\Gamma(n)=\gamma_{eff}(N_n+1)
\end{equation}
with $\gamma_{eff}=2\gamma_{eg} N_{ex} (\hbar\omega_e/k_BT)^3$.
In other words, if the previous conditions are satisfied, all the levels
of the ground state trap are equally loaded.

In our numerical simulations we consider
a $|g\rangle$ trap which is cut at an energy shell  $m_{max}=10$--$60$
(which for $\omega_g=2\pi\times 1$kHz, implies a trap depth of
$0.5$--$3.0 \mu$K).
In the simulations we have used a ``virtual'' trap of
$m_{max}+10$ energy shells
in order to take into account the atoms which do not decay into the trap,
and also to calculate the evaporative cooling dynamics.
We have checked that such chosen ``virtual'' trap does not affect the
physics of the problem. The levels $m>m_{max}$ are continuously emptied
in the simulations, and those atoms are considered as lost.

\begin{figure}[ht]
\begin{center}\
\epsfxsize=5.0cm
\hspace{0mm}
\psfig{file=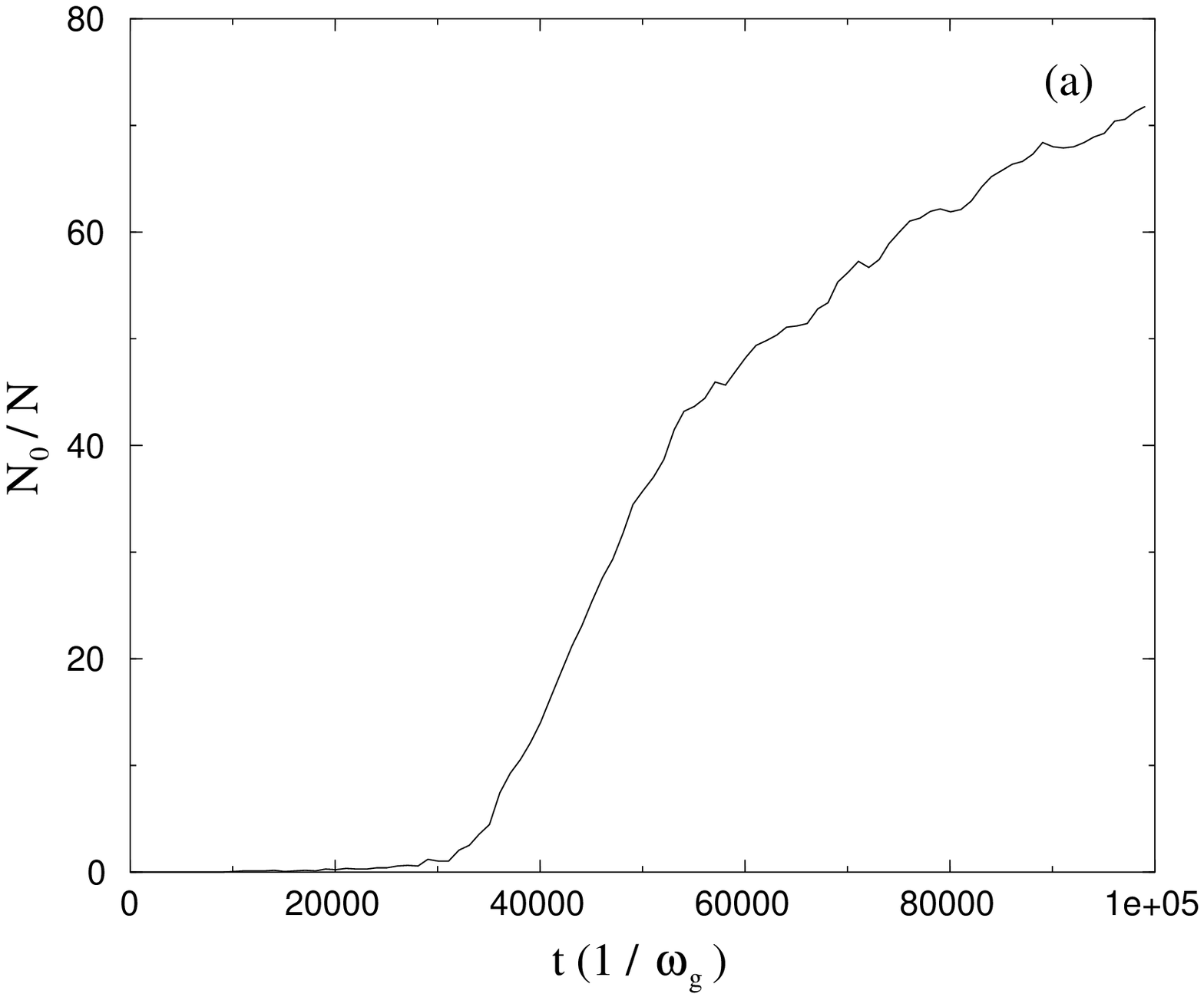,width=5.0cm} \\
\psfig{file=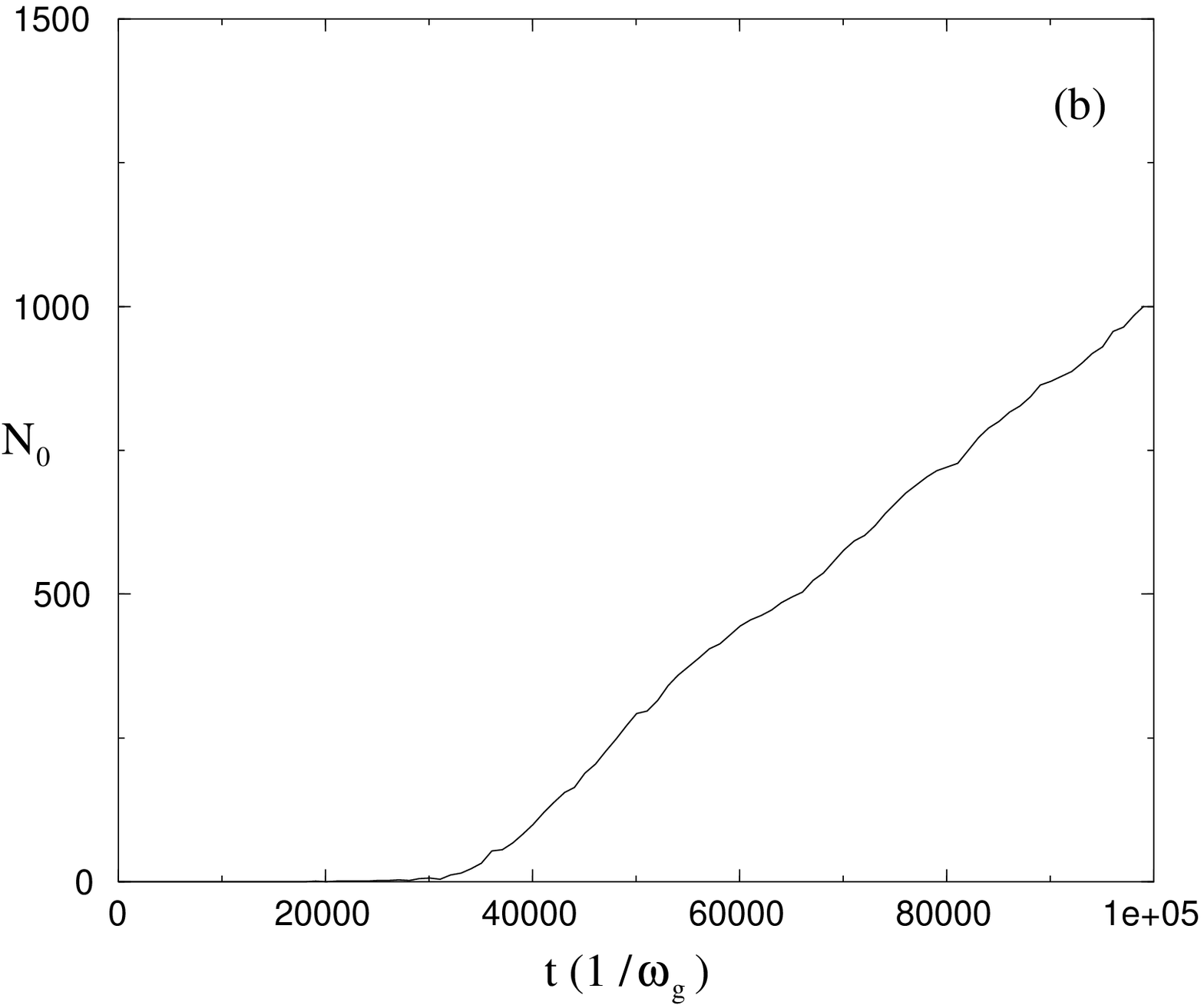,width=5.0cm} \\
\psfig{file=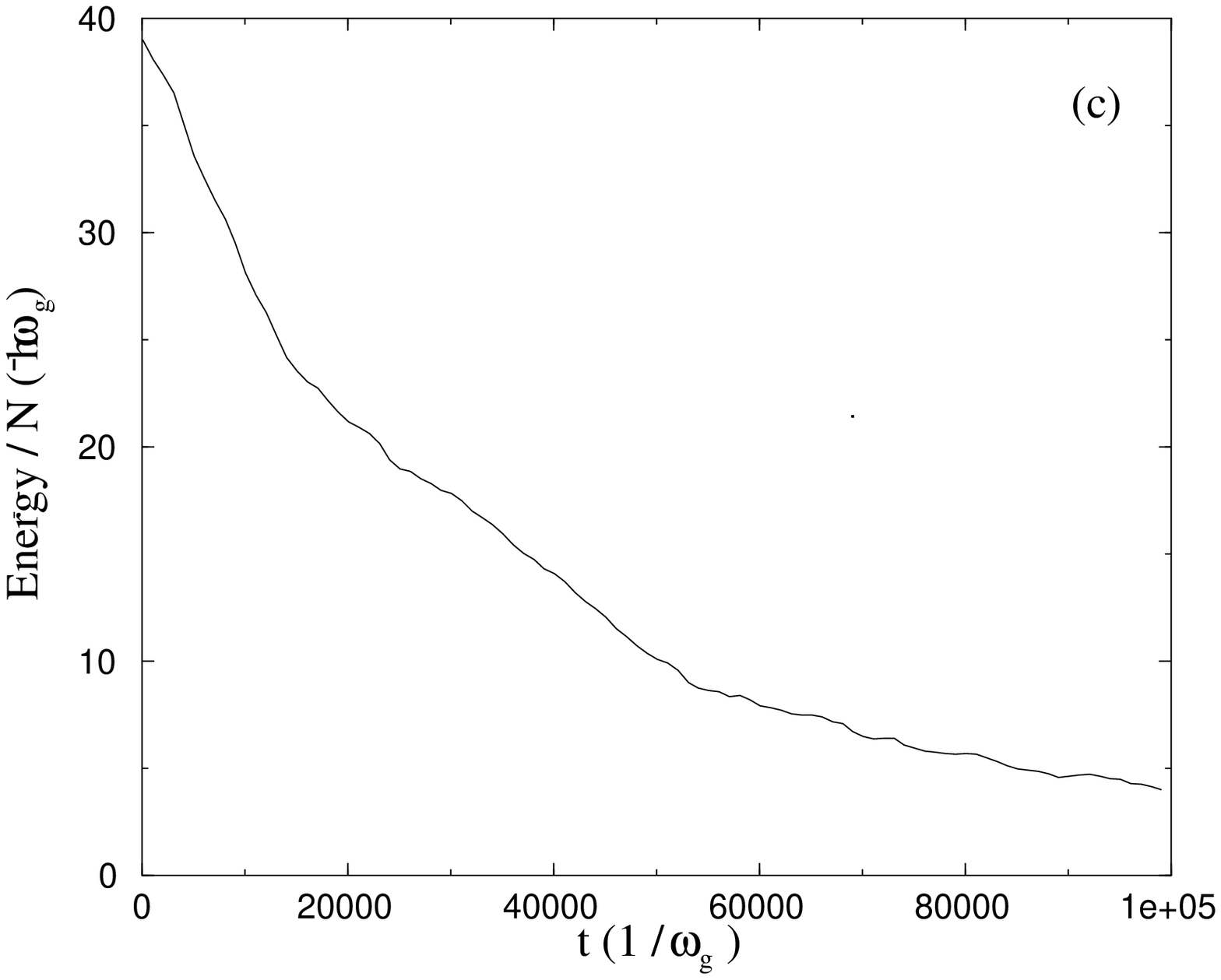,width=5.0cm} \\[0.2cm]
\caption{(a) Condensate fraction ($N_0/N$) as a function of time, in units
of the $\omega_g^{-1}$, for the case of $\gamma_{eff}=
0.01\omega_g$, $a=6$nm and $m_{max}=50$; (b) Dynamics of $N_0$ for the same situation;
(c) Evolution of the energy per particle for the same conditions.}
\label{fig:3}
\end{center}
\end{figure}
We have numerically simulated the loading dynamics of an initially empty
$|g\rangle$ trap for different values of the parameter $\gamma_{eff}$,
$m_{max}$ and the scattering length.
It is worthy to note that $\gamma_{eff}$ is related to the phase
space density of the thermal cloud
by the relation
$\gamma_{eff}=2\gamma_{eg}\times 5.2 n_{ex}\lambda(T)^3$, being
$\lambda(T)=(2\pi\hbar^2/Mk_BT)^{1/2}$ the thermal de Broglie wavelength, and
$n_{ex}$ the density in the $|e\rangle$ trap. Since the $|e\rangle$ atoms are
considered at a temperature far above the critical temperature of the onset
of the condensation, then $n_{ex}\lambda(T)^3$ is necessarly much smaller
than
1, restricting the possible range of values that $\gamma_{eff}$ can take.
We study below the dependence of the loading dynamics on
$\gamma_{eff}$. It is also interesting to point out the limits of the
large--temperature approximation, in which all the levels are equally
loaded. If such approximation is valid,
for a fixed phase space density $\phi$, a constraint to the density of
the reservoir is introduced by
$n_{ex}\gg\phi (M\omega\tilde m_{max}/2\pi\hbar)^{3/2}$, with
$\tilde m_{max}=m_{max}+\omega_{rec}/\omega_g$. For the case of $^{52}$Cr and
$\omega_g=2\pi\times 1$kHz, this
constraint implies $n\gg 7.45\phi m_{max}^{3/2}\times10^{11}$cm$^{-3}$.

As an example of loading dynamics we show in
Fig.\ \ref{fig:3} the case of $\gamma_{eff}=0.01\omega_g$,
$m_{max}=50$, and $\omega_g=2\pi\times 1$kHz.
For the case of $^{52}$Cr, $2\gamma_{eg}=200s^{-1}$, this would
imply a phase space density $\phi=10^{-5}$, and a constraint
$n_{ex}\gg 10^{8}cm^{-3}$
in order to satisfy the large--temperature approximation.
Due to the random character of the process, we have averaged over several
Monte Carlo realizations, in order to obtain smoother curves.
Typically the loading process is characterized by a time scale
of the onset of condensation,  after which, as observed in Fig.\ \ref{fig:3},
 the condensation
in the ground--state of the trap appears.
In the case of Fig. \ref{fig:3} the onset of the condensation appears
at approximately $3\times 10^4 \omega_g^{-1}$, which for
$\omega_g=2\pi\times 1$kHz would require $t=4.8$s.
After a time $10^5 \omega_g$ ($16$s) more than $1000$ atoms are
condensed with a condensate fraction of $70\%$.

Let us discuss the physics behind the presented numerical results.
Since the trap is initially empty, and all the loading rates are equal,
the higher shells of the trap are initially
loaded with larger probability, due
to their larger degeneracy. Once a sufficient number of particles is
loaded inside the trap, the collisions produce the evaporation of part
of the particles, whereas part of them are transported via collisions
to lower levels of the trap. During this process the energy per particle is
continuously decreased in the trap (see Fig.\ \ref{fig:3}(c)).
Finally, the condensation is produced
in the ground state. Once this happens,
a new mechanism which reduces the energy per particle in the trap appears,
namely the bosonic enhancement of the
spontaneous emission into the condensate. In effect, the condensate
loading becomes faster. When the number of condensed particles
becomes large the QBME equation is no more valid, as
discussed in Sec.\ \ref{sec:colis}. The mean field effects
appearing beyond the weak condensation
should not, however,  distort the qualitative effect of speeding up of
the loading rate\cite{idzia}.
It is also worthy to note that for condensate densities
larger than $10^{15}$ atoms/cm$^3$ inelastic processes (as three--body
recombination) are expected. Such processes would lead to
losses of condensed atoms, which can be repaired
by the continuous loading in a similar way as
described below for the case of atom laser outcoupling.
Let us also note that, eventually, if the number of condensed particles
were much larger
than the number of levels in the trap, the system would enter
into the so--called Bosonic Accumulation Regime, where the vast majority of
the decays were produced into the condensed state.

\begin{figure}[ht]
\begin{center}\
\epsfxsize=6.0cm
\hspace{0mm}
\psfig{file=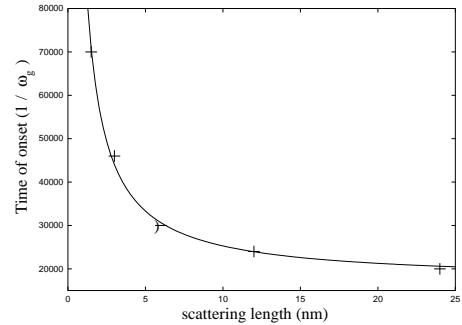,width=6.0cm} \\[0.3cm]
\caption{Time of onset of the condensation for $\gamma_{eff}=0.01\omega_g$,
$m_{max}=50$, as
a function of the scattering length.}
\label{fig:5}
\end{center}
\end{figure}

With this physical picture in mind, it is possible to understand
the dependences on the different physical parameters. From the
previous discussion it becomes clear the crucial role played by 
the collisions in the process. Therefore the larger the
collisional probability, the faster the condensation is achieved.
This  can be obtained in a two--fold way, (i) by increasing the
scattering length $a$ and/or (ii) by increasing the number of
trapped atoms. Point (i) have been studied in Fig.\ \ref{fig:5},
where we have analyzed the case of $m_{max}=50$,
$\gamma_{eff}=0.01\omega_g$ for values of the scattering length ranging
from $a=1.25$nm to $a=24$nm. As pointed out previously, larger
values of the scattering length produce faster condensation
onset. Note, however, that the time of onset of the condensation
reaches a constant value for large scattering length; this is
simply because there is, as pointed above, a time in which the
initially empty trap is being loaded, and in this initial time
the collisions play no significant role. Such initial time
depends basically on $m_{max}$ and $\gamma_{eff}$, as pointed out
below.

\begin{figure}[ht]
\begin{center}\
\epsfxsize=6.0cm
\hspace{0mm}
\psfig{file=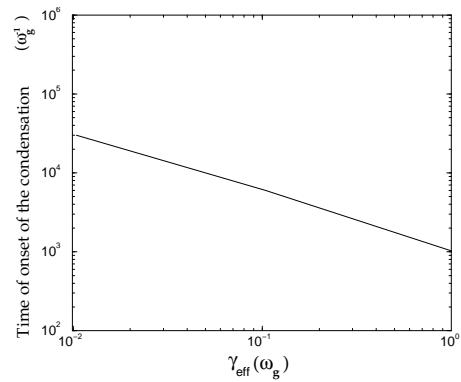,width=6.0cm} \\[0.3cm]
\caption{Time of onset of the condensation for $a=6$nm, $m_{max}=50$, as
a function of $\gamma_{eff}$.
}
\label{fig:4}
\end{center}
\end{figure}

\begin{figure}[ht]
\begin{center}\
\epsfxsize=6.0cm
\hspace{0mm}
\psfig{file=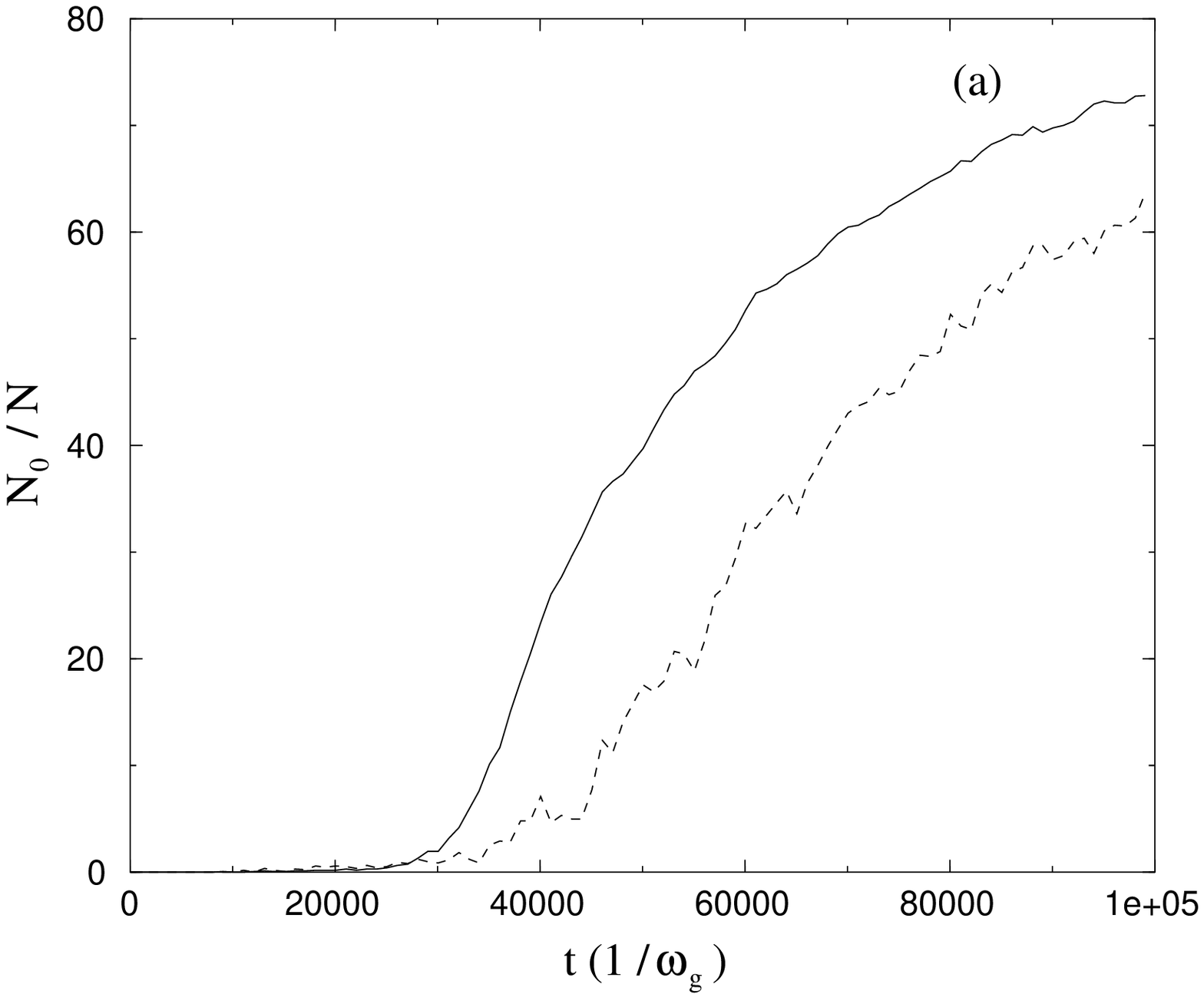,width=6.0cm} \\
\psfig{file=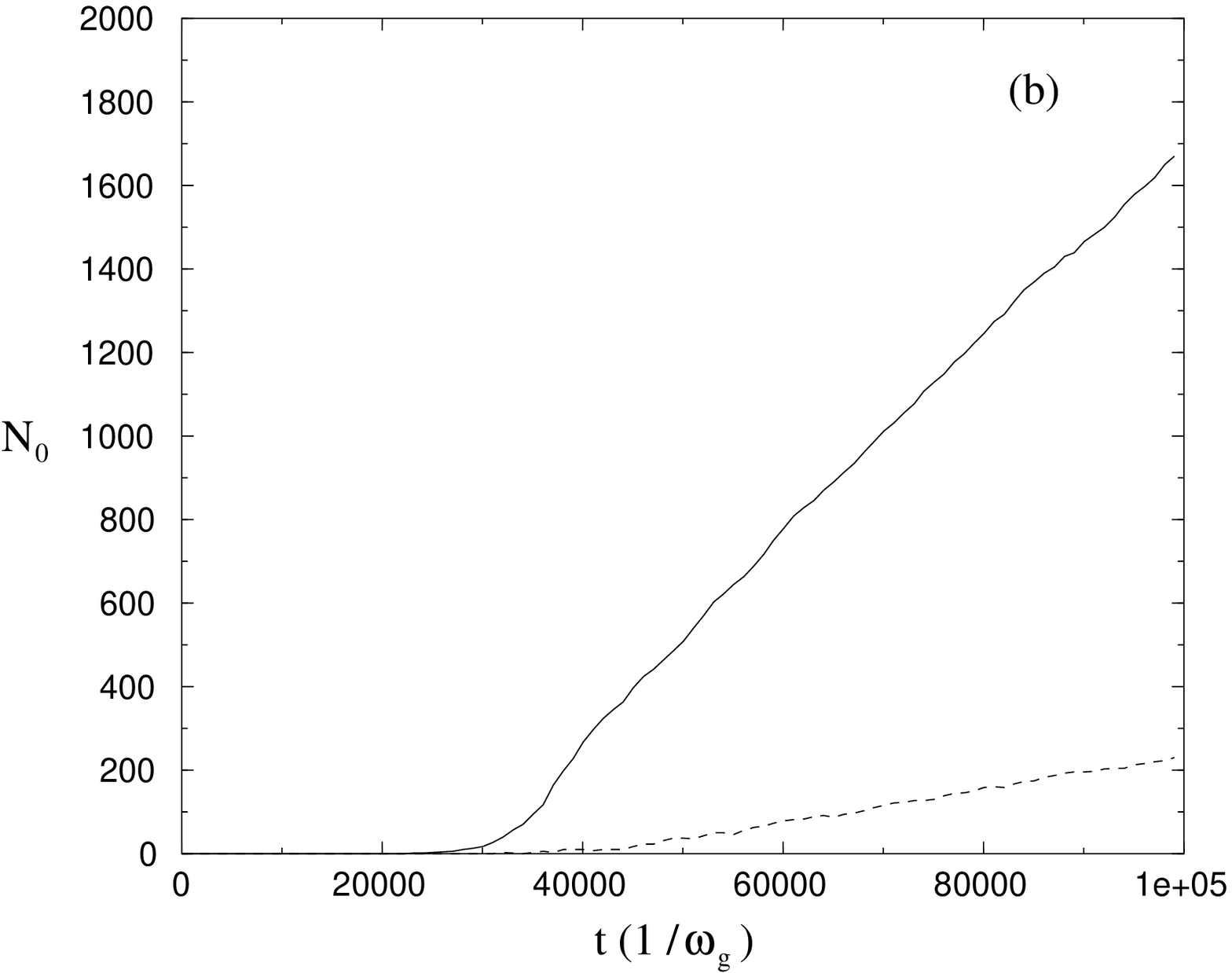,width=6.0cm} \\
\psfig{file=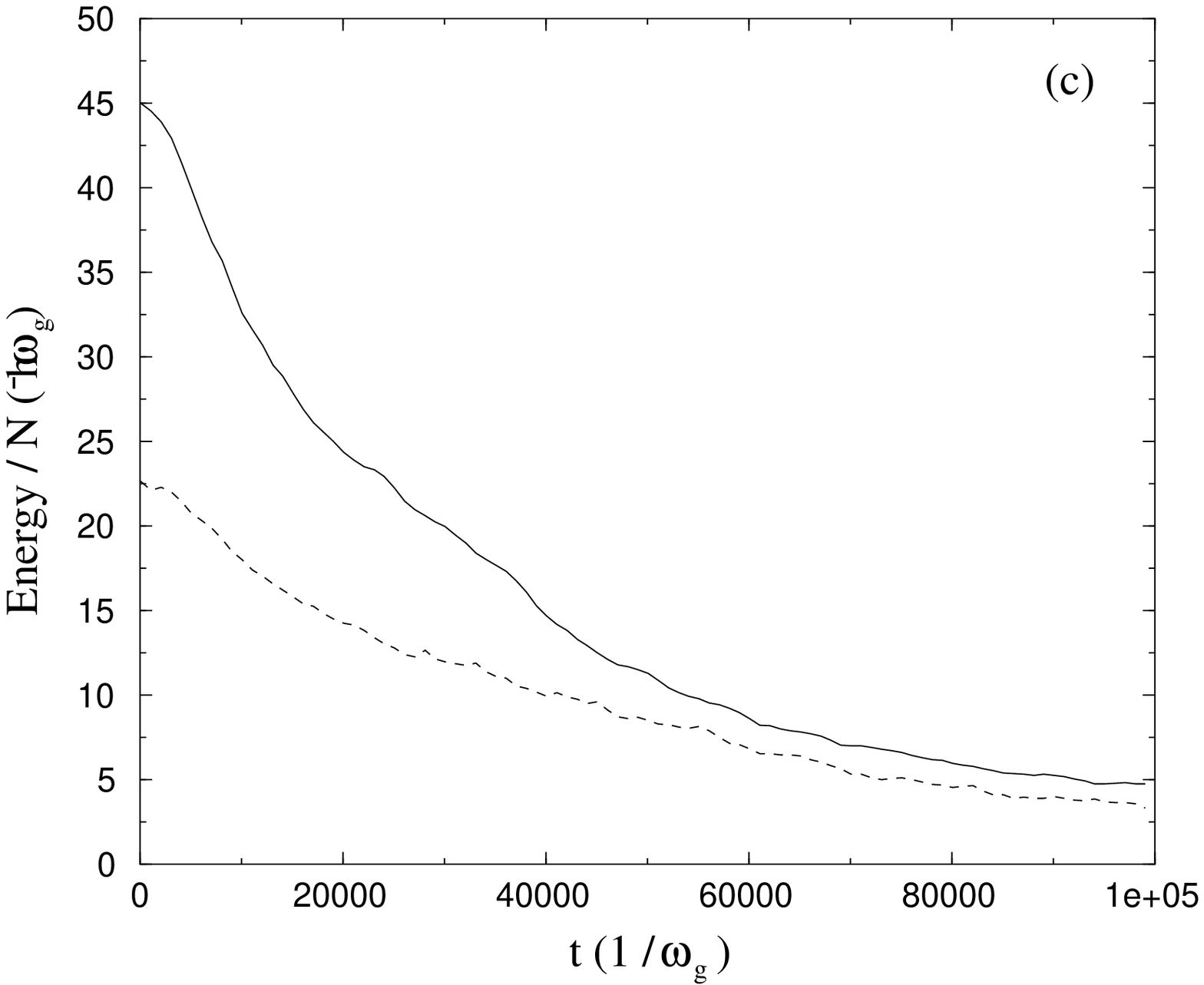,width=6.0cm} \\[0.3cm]
\caption{(a) Condensate fraction ($N_0/N$) as a function of time, in units
of $\omega_g^{-1}$, for the case of $\gamma_{eff}=
0.01\omega_g$, and $a=6$nm, and $m_{max}=60$ (solid) and $30$ (dashed); 
(b) Dynamics of $N_0$ for the same situation;
(c) Evolution of the energy per particle for the same conditions.}
\label{fig:6}
\end{center}
\end{figure}

A faster increase of the number of particles in the trap can be
achieved in two different ways:
\begin{itemize}
\item Increasing $\gamma_{eff}$.
In Fig.\ \ref{fig:4} we show the example of $m_{max}=50$, and a
scattering length $a=6$nm, for $\gamma_{eff}/\omega_g$ ranging between $1$
and $0.01$. As one could expect the onset of the condensation
appears faster for faster loading. In order to achieve a larger
$\gamma_{eff}$, it is necessary either to increase the phase
space density of the thermal cloud, or perhaps more
interestingly to consider experimental situations with bigger
$\gamma_{eg}$. Note that an increasement of $\gamma_{eg}$ by a
factor $100$ in the case consider above, still very well
satisfies the requirements of the BRE. We should stress once more
at this point that one of the main advantages of the BRE is the
fact that contrary to Festina Lente it is possible to pump with a
frequency larger than that of the loaded trap, without having
problems with the photon reabsorption. Therefore, it could be in
principle possible to consider faster natural transitions, or
even controllable quenched ones. In such cases, a fast onset of the
condensation could be possible for low phase space densities of
the reservoir.

\item Increasing $m_{max}$. In this way the total number of atoms in the trap
becomes larger at any given time.
As an example, we have considered in Fig.\ \ref{fig:6} the case
of $\gamma_{eff}=0.01\omega_g$, for
$m_{max}=30$ and $60$.
The increasement of the condensate fraction becomes faster for 
larger $m_{max}$, although the onset time remains approximately the same 
(Fig.\ \ref{fig:6}(a)).
The absolute number of
condensed atoms
is larger for larger values of $m_{max}$
(Fig.\ \ref{fig:6}(b)),
because the total number of atoms is also larger.
Therefore for practical purposes (e.g. detection) it would be
recommendable to use larger $m_{max}$.
However, we must stress at this
point that a larger $m_{max}$ needs a larger reservoir density
$n_{ex}$ if one wants to remain  within the limits of
the large--temperature approximation.

\end{itemize}

\begin{figure}[ht]
\begin{center}\
\epsfxsize=6.0cm
\hspace{0mm}
\psfig{file=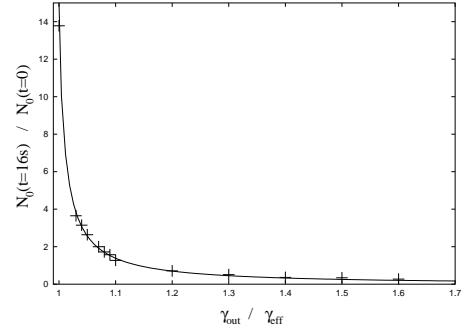,width=6.0cm} \\[0.3cm]
\caption{Number of condensed particles after $16$s
(after initially loading an empty $|g\rangle$ trap during 
$350$ms) 
for the case of $\omega=2\pi\times 1$kHz, $m_{max}=10$, 
$\gamma_{eff}=6.28\omega_g$, and different outcoupling rates $\gamma_{out}$.}
\label{fig:7}
\end{center}
\end{figure}

We have also analyzed the case in which an already formed
condensate is emptied via outcoupling, and continuously pumped
via spontaneous emission from the thermal reservoir. 
In our simulations we just consider outcoupling from the 
condensate, although similar methods could be employed 
to simulate losses affecting the whole trap. 
We simulate without outcoupling
the creation of a condensate as described above, for the case of
$\gamma_{eff}=6.28\omega_g$ , and $m_{max}=10$. For the case of
$^{52}$Cr, and $2\gamma_{eg}=200$s$^{-1}$. This represents a
quite large phase space density $6\times 10^{-3}$; however, we
must again stress that the BRE allows to work with much larger
$\gamma_{eg}$, and therefore with much lower phase space
densities of the reservoir. At $t=350$ms (when $N_0\simeq 950$),
we begin the outcoupling. We have analyzed different outcoupling 
rates $\gamma_{out}$ (Fig.\ \ref{fig:7}), and monitored the 
population $N_0$ after $16$s. 
This allows us to find a critical 
threshold $\xi_0$ (for this case $1.14$)
for the ratio $\xi=\gamma_{out}/\gamma_{eff}$. 
For $\xi>\xi_0$ the loading (``gain'') is faster than the 
outcoupling (``loss''), and the number of condensate atoms 
increases with time. For $\xi<\xi_0$, the number of particles decreases 
and stabilizes for a lower $N_0$. For $\xi\ll\xi_0$ no condensate 
can be kept.

\begin{figure}[ht]
\begin{center}\
\epsfxsize=6.0cm
\hspace{0mm}
\psfig{file=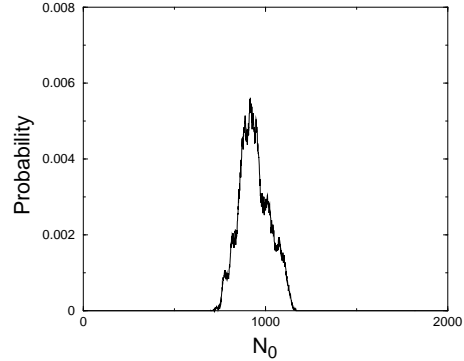,width=6.0cm} \\[0.3cm]
\caption{Condensate population, averaged during $40$s
for the case of $m_{max}=10$, $\gamma_{eff}=6.28\omega_g$,
$\omega_g=2\pi\times 1$kHz, $\gamma_{out}=(1.17-f(t))\gamma_{eff}$,
with $0<f(t)<0.05$ chosen randomly from an uniform distribution.}
\label{fig:8}
\end{center}
\end{figure}

It turns out to be important to maintain the population
as constant as possible, for the reasons that we clarify below, 
and therefore to work in the regime of $\xi=\xi_0$.
It is however not an easy task, due to the stochastic nature of
both, the collisions and the pumping mechanism. In order to
stabilize the noise optimally, i.e. to preserve the population of
the condensate as constant as possible it is useful to introduce
a random temporal variation of the outcoupling rate. Fig.\
\ref{fig:8} shows the averaged distribution of population of the
condensate during $40$ s of continuous outcoupling, for the case
of an outcoupling rate $\gamma_{out}=(1.17-f(t))\gamma_{eff}$,
with $0<f(t)<0.05$ chosen randomly from an uniform distribution, 
and the same conditions as in Fig.\ \ref{fig:7}.
The population of the condensate is maintained quasiconstant with
an average value of $\langle N_0\rangle=940$, and a variance
$(\langle N_0^2\rangle-\langle N_0\rangle ^2)^{1/2}=80$. During
these $40$s, $3\times 10^5$ atoms are extracted from the
condensate, with a rate of $7500$ atoms/s.

Let us briefly comment about the importance of keeping the
population of the condensate as constant as possible. The
mean--field interaction translates the variations of the
condensate population into variations of the energy of the
outcoupled atoms, being the variance of the energy related to the
variance of the condensate density: $\sigma
(E)=4\pi\hbar^2a\sigma (n_0)/m$. Therefore, the narrower the
population distribution of the condensate, the more
"monochromatic'' will be the atom--laser source, and consequently
the larger the coherence time will be  \cite{Cirac}. Let us point
out finally, that an additional way to control the
fluctuations of the condensate population could be provided by
monitoring the energy of the outcoupled atoms, which would inform
about the variations of the condensate density. Such information
could be used in a feedback loop to dynamically adapt the
outcoupling rate to reduce the energy variance of the outcoupled
atoms, and therefore increase their temporal coherence.

\section{Conclusions}
\label{sec:conclus}

In this paper we have analyzed a possible mechanism which could allow
the creation and continuous loading of a condensate from a thermal
reservoir, by optical pumping.
In order to achieve such loading mechanism, it is necessary to guarantee
that the reabsorptions of the
spontaneously emitted photons do not lead to undesired heating of the atoms
in the trap. We have analyzed a particular
scheme which allows to satisfy such condition. In this scheme an atom
forms a three level $\Lambda$ system, in which one of the
transitions decays much faster than the other one. By using quantum
Master Equation techniques we have shown that the very small
branching ratio between both transitions induces very large reduction
of probability of the reabsorption processes
 which change the population in the lowest
state of the slower transition. We have explained such effect
by identifying the photon reabsorption as a process
whose probability depends on
the correlation between the reabsorption amplitudes at different times.
Such correlation is rapidly destroyed by the fast decay into the other
possible channel. The destruction of this correlation
causes the desired effect, i.e. the reduction of the
 ``bad'' reabsorption processes, responsible for possible heating.

Once we have shown that the reabsorption has
 no significative effect on the system,
we have analyzed the loading dynamics from a thermal reservoir, using
Monte Carlo simulations, including the atom--atom collisions in
the QBME formalism. We have analyzed the loading of an initially empty trap,
demonstrating that the onset of the condensation appears after
 a finite time, which depends on the physical parameters
of the system. The condensation appears
due to the joint combination of thermalization via
collisions, evaporative cooling due to the finite depth of the
considered trap, and bosonic enhancement of the pumping process.
We have also analyzed the
continuous refilling of the condensate, once it has been formed, taking
at the same time into account
continuously outcoupling. We have shown that the refilling mechanism
allows the compensation of the losses introduced by the outcoupling, and
we have analyzed the best strategies to keep the condensate population
quasiconstant, which is important in order to achieve a
``monochromatic'' atom laser output.
In the paper we have only analyzed the outcoupling mechanism, but
the same reasonings applies to possible condensate losses, produced
by inelastic processes, such as three--body recombination, or collisions
with the thermal atoms in the reservoir.

All our simulations and estimates have been done for Chromium atoms, and for
the parameters of the experiment currently performed at the University of
Stuttgart. It is however interesting to stress
that the same scheme is general, and in particular can be
applied for other atomic systems, such as Magnesium\cite{rasel}.
As a final remark, we would like
to stress that the mechanism of  avoiding  the ``bad'' reabsorption
processes, considered in this paper (i.e. regime of BRE)
 allows for  faster pumping than other reabsorption remedies
(such as Festina Lente, for instance),
and therefore allows for more effective compensation
of the condensate losses. It offers a novel and
 interesting perspective towards
a continuously loaded atom laser.

We acknowledge  support from Deutsche Forschungsgemeinschaft (SFB
407), from the EU through the TMR network ERBXTCT96-0002, and from
ESF PESC Programm BEC2000+.
We acknowledge fruitful
discussions with W. Ertmer, E. Rasel, K. Sengstock,
and M. Wilkens.


\begin{references}
\bibitem{foot2} The large--temperature approximation is assumed since it simplifies
the numerical simulation.
Although this approximation has certain physical consequences, we want to stress that
similar effects as the ones discussed in this paper are not only possible for
the case of lower temperatures, but should in principle allow for faster loading.
In such a case, the Franck--Condon factors
of the different transitions should be properly calculated (c.f. \cite{Martin,Santos}).
\end{references}
\end{document}